\documentclass[
aps,
pra,
superscriptaddress,
12pt,
reprint,
longbibliography,
nofootinbib,
floatfix
]{revtex4-1}

\usepackage[utf8]{inputenc}
\usepackage[T1]{fontenc}
\usepackage[english]{babel}

\usepackage{graphicx}

\usepackage{amsmath}
\usepackage{amssymb}
\usepackage{mathtools}

\usepackage[colorlinks=true, allcolors=blue]{hyperref}
\usepackage[all]{hypcap}  

\usepackage{braket}
\usepackage{xcolor}
\usepackage[normalem]{ulem}

\renewcommand{\vec}[1]{\boldsymbol{#1}}

\renewcommand{\d}{\partial}
\newcommand{\D}[1]{\beta_{#1}}

\begin{document}
	
	\title{Kato's theorem and ultralong-range Rydberg molecules}
	\date{\today}
	
	\author{Matthew T. Eiles}
	\email{meiles@pks.mpg.de}
	\affiliation{Max Planck Institute for the Physics of Complex Systems, 01187 Dresden, Germany}
	\author{Frederic Hummel}
        \thanks{Present address: Atom Computing, Inc., Berkeley, CA 94710, USA}
        \email{frederic@atom-computing.com}
	\affiliation{Max Planck Institute for the Physics of Complex Systems, 01187 Dresden, Germany}
	\begin{abstract}
		We consider non-adiabatic coupling in the  "trilobite"-like long-range Rydberg molecules created by perturbing degenerate high-$\ell$ Rydberg states with a ground-state atom. 
  Due to the flexibility granted by the high Rydberg level density, the avoided crossings between relevant potential energy curves can become extremely narrow, leading to highly singular non-adiabatic coupling. 
  We find that the gap between the trilobite potential curve and neighboring "butterfly" or "dragonfly" potential curves can even vanish, as in a conical intersection, if the gap closes at an internuclear distance which matches a node of the $s$-wave radial wave function. 
  This is an unanticipated outcome of Kato's theorem.
  
	\end{abstract}
	
	\maketitle
		\textit{Introduction - }Non-adiabatic physics in the context of ultracold Rydberg atoms has garnered new interest in recent years. 
 In long-range Rydberg molecules, the coupling between potential wells capable of supporting vibrational levels and dissociative potential curves has been investigated as a possible decay mechanism \cite{Duspayev2022,duspayev2022NJP,deiglmayr2016long}.
Theoretical and experimental work has shown that non-adiabatic coupling can be strong enough to induce non-perturbative shifts in vibrational binding energies \cite{Hollerith2019,Srikumar2023, Hummel2022}.
 In cold or ultracold Rydberg collisions, the available chemical reaction pathways for molecular formation, state-changing collisions, or ionization are often determined by the strength of non-adiabatic 
 coupling parameters \cite{Hummel2021prl,Schlagmuller2016x,Geppert2020,Niederprum2015}.
 Interacting Rydberg ions \cite{Gambetta2021,Magoni2023} and Rydberg aggregates \cite{Wuster2011,leonhardt2014switching,pant2021nonadiabatic,leonhardt2016orthogonal} have been proposed as systems with which to probe and control dynamics through conical intersections. 

 A long-range Rydberg molecule
 consists of a Rydberg atom with principle quantum number $n$ and a distant (located $R\sim n^2\,\text{a}_0$ away) "perturber" atom in its electronic ground state. 
Non-adiabatic physics are a particularly interesting aspect of this system due to the close connection between the potential energy curves and the Rydberg wave functions \cite{Greene2000,Shaffer2018,Fey2019rev,Eiles2019}.
The large size of the molecules makes them an ideal laboratory to explore beyond Born-Oppenheimer physics on exaggerated scales, and the flexibility provided by Rydberg state parameters allows for controllable enhancement or suppression of non-adiabatic effects and the possibility to steer ultracold chemical reactions.  
 
 In this article, we show how the nodal lines of the Rydberg wave functions can be linked to very strong, even singular, vibronic coupling between the "high-$\ell$" or "trilobite"-like states of a long-range Rydberg molecule. This effort extends previous work \cite{Hummel2021prl} which showed that singular non-adiabatic coupling can arise in apparent contradiction of the von Neumann-Wigner no-crossing rule \cite{vonNeumann1929}. 
 We make an unexpected connection between these conical intersections and a fundamental property of the Coulomb potential, Kato's theorem \cite{Kato1957} and its spatially-dependent generalization due to March \cite{March1986}.  These relate the total electron density to its $s$-state contribution alone.
	
	\textit{Theory - }The interaction between the two atoms is mediated by the rapidly moving Rydberg electron (at position $\vec r$), which only encounters the short-ranged forces from the perturber (at position $\vec R$) inside of a small volume centered on it.
	Across this region, the Coulomb potential is essentially flat, and 
	the potential $\hat V(\vec r,\vec R)$  is spherically symmetric with respect to the perturber. 
 It is therefore conveniently described by an expansion into partial waves $L$ defined with respect to the perturber. 
	The $S$-wave contribution to this interaction is the well-known Fermi pseudopotential \cite{Fermi1934}. 
	The contribution of each partial wave is determined by the electron-perturber scattering phase shift, $\delta_L(k)$, which depends on the internuclear distance through the semiclassical momentum  $k\equiv k(\vec R,n) = \sqrt{2/|\vec R|-1/n^2}$.  
	These phase shifts scale as $\delta_L(k)\sim \frac{k^2}{L^3}$ for $L\ge 2$ and, 
	since $k\ll 1$, the importance of higher-order partial waves decrease rapidly \cite{Omont1977,Hamilton2002,Chibisov2002}.
	Hence, in the general theory developed below for the adiabatic potential energy curves $U_K(\vec R)$, we give particular expressions only for the $N=3$ most relevant partial waves, $L = 0,1,2$ \cite{Giannakeas2020b}. The molecular states are denoted "trilobite", "butterfly", and "dragonfly", respectively \cite{Hamilton2002,Giannakeas2020b}. 
	To keep the algebra transparent, we assume $0<\delta_L(k)<\pi$.\footnote{In all alkali atoms, this condition holds everywhere except small $R$.  This assumption can be relaxed at the expense of more careful algebra involving the imaginary square roots in the $\beta_L$ terms defined in Appendix \ref{app:Wterms}. }
	
	We obtain the set of adiabatic potentials $U_K(\vec R)$ by solving the electronic Schrödinger equation
	\begin{equation}
		\label{eq:BO}
		\left[\hat H_e(\vec r)+\hat{V}(\vec r,\vec R)\right]{\psi_K(\vec r;\vec R)} = U_K(\vec R){\psi_K(\vec r;\vec R)}. 
	\end{equation}
	Although the interaction potential $\hat V(\vec r, \vec R)$ is described using electronic partial waves $L$ defined with respect to the perturber, a natural basis to expand $\psi_K(\vec r;\vec R)$ into is the eigenstates of $\hat H_e(\vec r)$. These are the Rydberg states $\phi_{n\ell m}(\vec r) = \frac{u_{n\ell}(r)}{r}Y_{\ell m}(\hat r)$ with angular momentum $0\le \ell \le n-1$ relative to the ionic core. As the diatomic system possesses cylindrical symmetry, each $m$ can be treated individually and we consider just $m=0$ below.

	\begin{figure*}
		\label{fig:cross}
		\centering
		\includegraphics[width=\textwidth]{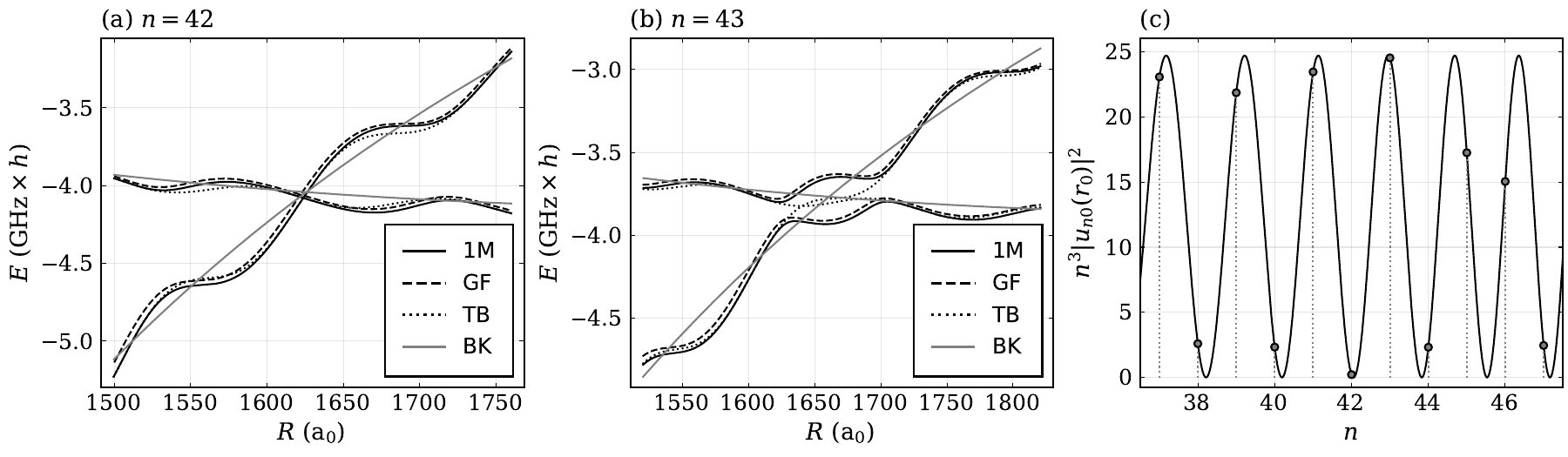}
		\caption[Potential crossings]{
			Crossings of the adiabatic potential energy curves at principal quantum numbers $n=42$ (a) and $n=43$ (b) calculated using different methods.
			1M: numerical eigenvalues of $V_{ll'}$ using one manifold of Rydberg states; GF: Green's function method; TB: the diagonal "trilobite" and "butterfly" potentials $U_S(R)$ and $U_P(R)$, respectively; and BK: the semiclassical Borodin-Kazansky model of Eq.~\ref{eq:BK}.  
			Energies are measured with respect to $-1/(2n^2)$. 
            (c) Nodes of the Rydberg $\ell=0$ orbital. The plotted curve is $n^3|u_{n0}(r_0)|^2$, where $r_0$ is obtained by solving $k^2=2/r_0 - 1/n^2$ for the $k$ value where $\delta_S(k) = \delta_P(k)$. When this function is zero, a conical intersection occurs. Narrow avoided crossings appear when a conical intersection occurs very close to an integer $n$.  }
	\end{figure*}
	
	As our focus here lies on the high-$\ell$ states, we neglect the quantum defects $\mu_\ell = \delta_\ell/\pi$ caused by deviations from a pure Coulomb potential in a non-hydrogenic atom.
We additionally neglect coupling to additional $n$ levels. 
	Both of these assumptions are well-justified here \cite{Eiles2023, Eiles2019},
 and permit the replacement of $\hat H_e(\vec r)$ by the number $-1/(2n^2)$. 
	The adiabatic potential curves $U_K(\vec R)$ are obtained by diagonalizing $\hat{V}$, whose matrix elements  in the degenerate $\ell$ subspace of a given $n$ are\footnote{Throughout, although in principle all variables defined here depend on $n$, we keep this dependence implicit except at the level of the hydrogen wave functions or energies.}
	\begin{equation}
		\label{eq:V}
		V_{\ell \ell'} = -\sum_{L=0}^{N-1} W_{\ell L}^\dagger W_{L\ell'}.
	\end{equation}
	$V_{\ell \ell'}$ shows how the interaction with the perturber causes an incoming Rydberg electron with angular momentum $\ell$ (relative to the ionic core) to scatter, via each partial wave $L$ (relative to the perturber), into a state $\ell'$. 
	The first three rows of the rectangular matrix $W_{L\ell}$ are given in Appendix \ref{app:Wterms}.
 
\textit{Non-adiabatic coupling - }
The strength of the non-adiabatic coupling between adiabatic states $K$ and $K'$, quantified by the derivative coupling matrix $\langle\psi_K\ket{\partial_R\psi_{K'}}$, is inversely proportional to the energy gap $U_K(R) - U_{K'}(R)$. 
After diagonalizing $V$, regions in the potential curves where non-adiabatic coupling becomes large can be identified by searching for small gaps.
As pointed out in Ref.~\cite{Hummel2021prl}, these can become arbitrarily small when they occur at a discrete $n$ value close to the position of a conical intersection in the potential surfaces defined as functions of $R$ and $n$, where $n$ is taken to be a \textit{continuous variable}. 

To predict the positions $(R_0,n_0)$ of such conical intersections it proves essential to represent the interaction operator in a different basis, namely the \emph{perturber spherical basis} composed of the non-orthogonal states 
	\begin{equation}
		\tilde \psi_L(\vec r;\vec R) = \sum_{L'}\left[\hat W\hat W^\dagger\right]^{-1/2}_{LL'}\sum_\ell W_{L'\ell}\phi_{n\ell 0}(\vec r).
	\end{equation}
 The transformation from the Rydberg basis to this one is accomplished using the left-inverse $\hat{S}$ satisfying
	\begin{align}
		\hat S&= (\hat{W}\hat{W}^\dagger)^{-1/2}\hat W,\quad \hat{S}\hat{S}^\dagger = \hat{1}_{N\times N}.
	\end{align}
	Using these definitions, it is straightforward to show that $\hat{V}=\hat{S}^\dagger\hat{S}\hat{V}\hat{S}^\dagger\hat{S}$ \cite{Eiles2023}, and thus  
	\begin{equation}
		\label{eq:BO2}
		\hat{S}(\vec R)\tilde{V}(\vec r,\vec R)\hat{S}^\dagger(\vec R)\tilde{\psi}_K(\vec r;\vec R) = U_K(\vec R)\tilde{\psi}_K(\vec r;\vec R),
	\end{equation}
	where $\tilde{\psi}_K\equiv \hat S\psi_K$ and $\tilde{V}\equiv \hat{S}\hat{V}\hat{S}^\dagger = -\hat{W}\hat{W}^\dagger$.
	The matrix element $\tilde{V}_{LL'}$ is proportional to the overlap $\langle \tilde \psi_L|\tilde\psi_{L'}\rangle$; these are computed explicitly in Appendix \ref{app:Wterms}. 
	Clearly, rather than dealing with the $n\times n$ matrix $\hat V$ of Eq.~\ref{eq:V}, it suffices to study the conditions necessary to obtain degenerate eigenvalues of the $N\times N$ matrix $\tilde V$. Semiclassically, the elements $\tilde V_{LL}$ are approximately \cite{Giannakeas2020a,Borodin1992}
	\begin{equation}
		\label{eq:BK}
		U_L(R) \sim -\frac{1}{2(n-\delta_L(k)/\pi)^2}.
	\end{equation}
\textit{Trilobite / Butterfly subspace - }We first consider the subspace with $L\le 1$, corresponding to the trilobite and butterfly states. 
 The $2\times 2$ matrix $\tilde V$ possesses degenerate eigenvalues if $\tilde V_{00}=\tilde V_{11}$ simultaneously as $\langle \tilde \psi_0|\tilde \psi_1\rangle= 0$. 
	That the latter condition can be met is \textit{a priori} not guaranteed: the overlaps determining the diagonal elements, for example, are nodeless (See Eqs. \ref{eq:q00} and \ref{eq:q11}). 
	However, employing the spatial generalization of Kato's cusp theorem \cite{Kato1957}, which was first derived and studied in the context of density functional theory \cite{Blinder1984, Shakeshaft1985, March1985, March1986} but also discovered in the calculation of electron-transfer in charged particle collisions \cite{Chibisov2000, Cherkani2001}, we obtain
	\begin{equation}
		\label{eq:kato}
		\langle \tilde \psi_0|\tilde \psi_1\rangle= -\left|\phi_{n00}(\vec R)\right|^2.
	\end{equation}
	This result -- that the coupling between trilobite and butterfly states is determined by the $s$-wave probability density alone -- shows that the coupling vanishes when 
 \begin{equation}
 \label{eq:swave}
     u_{n0}(R) = 0,
 \end{equation}
 and therefore degenerate eigenvalues are possible. Using the semiclassical result of Eq.~\ref{eq:BK}, we find the first condition, that the diagonal elements are equal, to occur when
 \begin{equation}
 \label{eq:phasematch}
     \delta_L(k) = \delta_{L'}(k)
 \end{equation}
for arbitrary $L$ and $L'$. 
 For $S$ and $P$ partial waves, if Eqs.~\ref{eq:kato} and \ref{eq:swave} hold at the same $(R,n)$ tuplet, 
 the two partial waves locally decouple and the potential surfaces will cross in a conical intersection. 
Remarkably, inserting Eq.~\ref{eq:swave} into the quantum formulas for the diagonal energies (Eqs.~\ref{eq:Vtilde}, \ref{eq:q00}, and \ref{eq:q11}),
	shows that Eq.~\ref{eq:phasematch} holds for the fully quantum calculation as well.  
   That the semiclassical condition perfectly matches the quantum one is another surprising conclusion stemming from Kato's theorem. 

   	Fig.~\ref{fig:cross} shows two extreme examples of the curve crossing between $S$ and $P$ states (for concreteness, we have taken Rb to be the perturber). For $n=42$ (Fig.~\ref{fig:cross}a) the $R$ value where Eq.~\ref{eq:phasematch} holds lies almost perfectly at a node of $u_{42,0}(R)$ (compare Fig.~\ref{fig:cross}c). 
    For $n=43$ (Fig.~\ref{fig:cross}b) this point lies nearly at an anti-node of the $n=43$ wave function.
    Therefore, the former case exhibits an extremely narrow crossing (on the sub MHz level) while the latter case possesses a pronounced avoided crossing. 
	In Fig.~\ref{fig:cross}(a,b) we compare four different calculations of the two potential curves. The curves labeled 1M result from the diagonalization of $\hat V$.  The curves labeled GF were obtained from a Green's function calculation \cite{Hamilton2002,Khuskivadze2002,Eiles2023GF}, which includes contributions from the entire Rydberg spectrum and not just a single Rydberg manifold as in the 1M case \cite{Fey2015}.
 These two complementary methods agree almost perfectly, especially regarding the existence and size of the narrow avoided crossing. 
 We also show two approximations, the so-called BK model \cite{Borodin1992}, equivalent to Eq.~\ref{eq:BK} but derived using a different approach, and the TB curves, which show the exact diagonal elements $\tilde{V}_{00}(R)$ and $\tilde{V}_{11}(R)$ defined in Eq.~\ref{eq:Vtilde} and below. 

\textit{Dragonfly contributions - }	We now include the effect of $L=2$ partial waves.  The coupling between $S$ and $D$,
	\begin{align}
 \label{eq:02coup}
		\langle \tilde \psi_0|\tilde \psi_2\rangle \propto\frac{u_{n0}(R)}{4\pi R^3}\left[2u_{n0}(R) - Ru_{n0}'(R)\right],
	\end{align}
	is also oscillatory. It vanishes when $u_{n0}(R) = 0$ or $\frac{d}{dR}\ln u_{n0}(R) = \frac{2}{R}$. When the former condition holds, the $S$ partial wave decouples from both $P$ and $D$ waves simultaneously. 
	Unfortunately, the algebra of  higher $L$ values becomes  tedious \cite{Omont1977,idziaszek2006pseudopotential}. 
 We speculate that the $S$-state decoupling persists even for higher $L$ values, but further effort is needed to make this generalization rigorous. 
A degeneracy in the $2\times 2$ subspace of $S$ and $D$ levels is also possible, occurring when the expression in Eq.~\ref{eq:02coup} vanishes simultaneously as $\delta_2(k) = \delta_0(k)$. This semiclassical condition, unlike the $S$$P$ case, coincides with the quantum condition only when $R\gg 1$ (Eq.~\ref{eq:bigR}). 

Surprisingly, the coupling between $P$ and $D$ states does not oscillate (Eq.~\ref{eq:q12}), and therefore these curves \textit{cannot} cross. It is intriguing that, just as Kato's theorem shows the special role played by the $s$-wave function (defined with the origin at the Rydberg ion), it shows how the $S$-wave molecular state, defined with the origin at the perturber, also behaves in a non-generic way. 

	\textit{Discussion - } We showed that the $L=0$ trilobite state decouples from the $L=1,2$ partial waves, and likely all higher partial waves, whenever Eq.~\ref{eq:swave} is satisfied. This is a direct result of Kato's theorem. It is particularly intriguing since the weight of the $\ell=0$ state in the wave functions $\ket{\tilde \psi_0}$ and $\ket{\tilde \psi_1}$ is almost negligible. 
	 If the $S$ and $P$-wave phase shifts are equal when Eq.~\ref{eq:swave} is satisfied, a conical intersection exists and its effects will emerge in the non-adiabatic coupling of nearby integer $n$ levels. Such a conical intersection cannot occur between butterfly and dragonfly potential curves. 
 
 It is interesting to contrast these results with what was observed in Ref.~\cite{Hummel2021prl} for the crossing of a trilobite state with a quantum defect state, with angular momentum $\ell_0$ and a non-zero quantum defect $\mu_{\ell_0}$ Semiclassically, the relevant potential curves become degenerate when 
	\begin{equation}
		\pi\mu_{\ell_0} = \delta_S(k),\label{eq:condi-qd}
	\end{equation}
 at the same $(R_0,n_0)$ as $u_{n_0\ell_0}(R_0) = 0$. 
 Eq.~\ref{eq:condi-qd}
	bears similarity to Eq.~\ref{eq:phasematch}, but now requires the phase accumulated by scattering off of the non-hydrogenic core of the Rydberg atom to match that accumulated from scattering off of the perturber. 
 This is closely connected 	
	with the fact that the $R$-dependent scattering phase shifts play the role of quantum defects in the Rydberg formula given by Eq.~\ref{eq:BK}. 
	The second condition is analogous to Eq.~\ref{eq:swave} but differs in a key way which again illustrates the counterintuitive message of Kato's theorem.
	For a quantum defect state, it is that radial wave function which must possess a node.  
 This carries a certain degree of physical intuition as this state is the dominant component of one of the electronic states  in the system. 
 On the other hand, for the trilobite and butterfly state interaction, there is nothing in the scattering problem or in the pure Coulomb interaction to single out a specific $\ell$. 
 However, because Kato's theorem places fundamental importance on $\ell=0$, this is the state which matters in the end, in what appears to be a surprising accident of the Coulomb potential.

	\appendix
	\section{Matrix elements of  $\tilde V$}
	\label{app:Wterms}
 The $L\le 2$ matrix elements of the $N\times n$ rectangular matrix $\hat{W}$ are
		\begin{subequations}
		\label{eq:W}
		\begin{align}
			W_{0\ell} &= \D{0}\,\phi_{n\ell0}(\vec{R}) \\
			W_{1\ell} &= \D{1}\frac{\d}{\d R}\phi_{n\ell0}(\vec{R})\\
			W_{2\ell} &= \D{2}\left(\frac{3}{2}\frac{\d^2}{\d R^2}+\frac{k^2}{2}\right)\phi_{n\ell0}(\vec{R}),
		\end{align}
	\end{subequations}
	where $\D{L} = \sqrt{2(2L+1)\pi k^{-(2L+1)}\tan\delta_L(k)}$.
	 The matrix elements of $\tilde V$ are
	\begin{equation}
		\label{eq:Vtilde}
		\tilde V_{LL'}= -\D{L}\D{L'}\tilde{Q}_{LL'},
	\end{equation}
	where 
	\begin{subequations}
		\label{eq:Qtilde}
		\begin{align}
			\tilde Q_{LL'}&= Q_{LL'},\ L, L'\le1,\\
			\tilde Q_{L2}&= \frac{1}{2} (3Q_{L2}+k^2Q_{L0}),\ L\le1,\\
			\tilde Q_{22} &= \frac{1}{4} (9Q_{22} + 6k^2Q_{20} + k^4Q_{00}). 
		\end{align}
	\end{subequations}
	These formulas make use of the quantity
	\begin{align}
		\label{eq:Q}
		Q_{\alpha\beta} &= \sum_{\ell=0}^{n-1} \sum_{m=-\ell}^{m=\ell} \frac{\d^\alpha}{\d R^\alpha} \phi_{n\ell m}^\ast(\vec{R}) \frac{\d^\beta}{\d R^\beta} \phi_{n\ell m}(\vec{R}).
	\end{align}

	\section{Computed Q values}
	\label{app:Qterms}
	In Appendix \ref{app:Wterms}, the matrix $\tilde V$ is defined in terms of the overlap matrix $\tilde Q$, which is in turn defined using $Q_{\alpha\beta}$ (Eq. \ref{eq:Q}) as a sum over the degenerate $\ell$ and $m$ states. 
 This summation can be performed analytically, as described in \cite{Chibisov2000,Cherkani2001,Eiles2019}. 
  In doing so, all $\tilde Q$ terms can be defined in terms of only the $s$-wave radial wave function and its derivative, as summarized below. We use $u\equiv u_{n0}(R)$ and $u'=u_{n0}'(R)$ to shorten the notation:
   \begin{align}
		\label{eq:q00}
		\tilde Q_{00} &= \frac{1}{4\pi} \left[k^2 u^2 +u'^2\right]\\
		\label{eq:q11}
		\tilde Q_{11} &= \frac{k^2}{3}Q_{00} -\frac{1}{6\pi R^2}\left(2uu'-\frac{u^2}{R}\right)\\
  \label{eq:q22}
  \tilde Q_{22}&=\frac{1}{20\pi n^2R^3}\Bigg[\frac{2n^4(2R+9)+R^3-4n^2R^2}{n^2}u'^2\nonumber\\&+\frac{4(n^2(2R-9)-R^2)}{R}uu'\\
	\nonumber&+4(2-3R)u^2\\&+\frac{6n^2R^4-R^5+(18+R(8R-7))n^6}{n^4R^2}u^2\Bigg].\nonumber
	\end{align}
For $R\gg 1$,
\begin{align}
\label{eq:bigR}
	\tilde Q_{22}&\sim\frac{k^2}{20\pi}\Bigg[(ku')^2+\frac{4}{R^2}uu'+(k^2u)^2\Bigg].
\end{align}
	We note here that the expressions given for $Q_{11}$ and a related quantity in Refs. \cite{Eiles2019} are incorrect. We report the correct formulas here for completeness:
		\begin{align}
			\Upsilon_{33}&=\frac{4\pi R^2k^2Q_{00}-uu'-u^2/R}{12\pi R^2}\\
			\Upsilon_{22}&=Q_{11}\nonumber\\
			&=\Upsilon_{33}-\frac{u_{n0}(R)}{4\pi R^3}(Ru'-u).
	\end{align}
 For large $n$ values the missing term does not lead to noticeable differences. 
The various off-diagonal couplings are
\begin{align}
    \tilde Q_{01} &= -|\phi_{n00}(R))|^2 = -\frac{u^2}{4\pi R^2}\\
		\tilde Q_{02}&=\frac{u}{4\pi R^3}\left[2u - Ru'\right]\\
  \label{eq:q12}\tilde Q_{12}&=-\frac{1}{8\pi R^2}\Bigg[3\left(\frac{u}{R}-u'\right)^2+k^2u^2\Bigg]. 
	\end{align}
For completeness, another useful result is
\begin{align}		Q_{02}&=-\frac{k^2}{3}Q_{00}+\frac{1}{6\pi R^2}\left(\frac{2u^2}{R}-uu'\right).
	\end{align}

	\section{Green's function}
		
The closed-form Coulomb Green's function \cite{Hostler1963} leads to a transcendental equation \cite{Hamilton2002,Eiles2023GF}
	\begin{align}
		\label{eq:GF}
		0 &= A_0(A_1-A_{01}),
  \end{align}
  where
  \begin{align}
      \label{eq:gfterms}
		A_0 &=1 - \Phi_v\frac{\tan\delta_0}{k}\\
		A_1&=1+\left(\Phi_{vvv} - 3\Phi_{uuv}\right)\frac{\tan\delta_1}{k^3}\\
		A_{01}&=\frac{\tan\delta_0/k}{1 - \Phi_v\tan\delta_0/k}3\Phi_{uv}^2\frac{\tan\delta_1}{k^3},
	\end{align}
	whose solutions give the potential energy curves $V(R)=-\frac{1}{2\nu(R)}^2$. 
	Here, $\nu(R)$ is the $R$-dependent principle quantum number whose non-integer part gives the deviation from the unperturbed hydrogen levels caused by the perturber. 
	The various $\Phi_x$ terms in Eq.~\ref{eq:GF} relate to derivatives of the Coulomb Green's function. 
	The zeros of $A_0$ and $A_1$, computed individually, give the $L=0$ and $L=1$ potential curves. 
 $A_{01}$ describes the coupling between these terms and vanishes whenever
	\begin{equation}
		0=\Phi_{uv}  = - \frac{\nu \Gamma(1-\nu)}{2R^{2}} M_{\nu,1/2}\left(\frac{2R}{\nu}\right) W_{\nu,1/2}\left(\frac{2R}{\nu}\right),
	\end{equation}
	where $M$ and $W$ are Whittaker functions. 
	When $\nu$ is an integer, the nodes of $M_{\nu,1/2}$ and $W_{\nu,1/2}$ coincide with those of $u_{n0}$, and hence we recover the result from diagonalization discussed in the main text. 

 \section{Rubidium phase shifts}
 The existence of narrow avoided crossings at a specific $n$ value does depend sensitively on the electronic phase shifts, and in particular on the energy where they are computed to become identical. For reproducibility, we give here the phase shifts used in our calculations, which match those of Ref.~\cite{Engel2019}.
	\begin{figure}[h]		\label{fig:phases}
		\centering
		\includegraphics[width=0.35\textwidth]{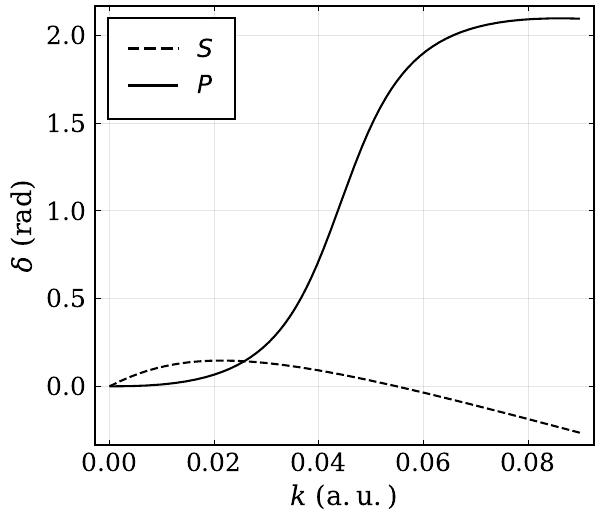}
		
		\caption{$S$- and $P$-wave phase shifts used to compute Fig.~\ref{fig:cross}, taken from \cite{Engel2019}.}
	\end{figure}
For other sets of phase shifts, whether they are computed using different methods or measured experimentally, we would expect the $n$ values where conical intersections nearly occur to differ due to any change in the intersection point. 
\end{document}